\UseRawInputEncoding
\documentclass[aps,onecolumn,10pt]{revtex4}
\usepackage{amsmath}
\usepackage{amssymb}
\usepackage{hyperref}
\hypersetup{colorlinks=true,linkcolor=red,citecolor=green,urlcolor=blue}
\numberwithin{equation}{section}
\numberwithin{equation}{section}

\begin{document}
\allowdisplaybreaks
\setcounter{equation}{0}

\title{Imprint of galactic rotation curves and metric fluctuations on the recombination era  anisotropy}

\author{Philip D. Mannheim}
\affiliation{Department of Physics, University of Connecticut, Storrs, CT 06269, USA \\
philip.mannheim@uconn.edu\\ }

\date{December 28 2022}

\begin{abstract}

In applications of the conformal gravity theory it has been shown that a scale of order 105 Mpc due to large scale inhomogeneities such as clusters of galaxies is imprinted on the rotation curves of galaxies. Here we show that this same scale is imprinted on recombination era anisotropies in the cosmic microwave background. We revisit an analysis due to Mannheim and Horne, to show that in the conformal gravity theory the length scale of metric signals that originate in the primordial nucleosynthesis era at $10^{9\circ}$K can fill out the entire recombination era sky. Similarly, the length scale of acoustic signals that originate at $10^{13\circ}$K can also fill out the entire recombination era sky. We show that the amplitudes of metric fluctuations that originate in the nucleosynthesis era can grow by a factor of $10^{12}$ at recombination, and by a factor of $10^{18}$ at the current time. In addition we find that without any period of exponential expansion a length scale as small as $10^{-33}$ cm can grow to the size of the recombination sky if it begins to grow at a temperature of order $10^{33}$ degrees. 
\end{abstract}

\maketitle

\section{Introduction}
\label{S1}

In the standard Newton-Einstein gravitational theory the local physics associated with the rotational properties of the stars and gas within individual spiral  galaxies and the global physics associated with cosmology are essentially decoupled. This decoupling originates in a decoupling theorem due to Newton that for a spherically symmetric distribution of matter and a $1/r$ potential, matter exterior to any spherical surface of interest does not contribute to the dynamics inside the surface. Thus with the rotation curves of spiral galaxies not showing the Kepler fall off associated with the luminous matter content of the galaxies, any attempt within Newtonian gravity to attribute this discrepancy to nonluminous dark matter would require that the dark matter be located within the galaxies of interest themselves. While dark matter does play a role in the formation of galaxies in the standard $\Lambda$CDM dark matter picture, any dark matter exterior to any individual galaxy does not affect the rotation curve of the material within that galaxy once its galactic halo is formed. Thus central to the dark matter program is that the structure of galactic rotation curves is determined solely by the local luminous and dark matter content within the given  galaxy and is independent of the global matter content of the rest of the universe.

This disconnect between local and global physics in standard dark matter theory (or at least in its formulation to date)  is exhibited in the fact that even though cosmological dark matter fluctuation theory can generate galactic dark matter haloes, the theory is currently unable to a priori determine how to match any of them with any given galactic luminous matter distribution.  For the moment dark matter theory is only able to fix otherwise free halo parameters (typically two for each galactic halo)  once a galactic rotation curve has been measured, and is not able to determine galactic rotation curves from a knowledge solely of the luminous matter content with any given galaxy. Thus for the 138 galactic sample that we discuss below dark matter theory currently needs 276 free parameters.

Not only is this shortcoming a challenge for dark matter theory, it turns out that some cosmological signatures have actually been detected in the systematics of the galactic rotation curves, signatures that can explicitly be associated with material outside of the galaxies.  Specifically, phenomenological evidence for such a connection was provided in \cite{Mannheim2011,Mannheim2012a,O'Brien2012}, where it was noted that for a broad sample of 138 spiral galaxies from bright spirals to low surface brightness galaxies and dwarfs the numerical values of the centripetal accelerations at the last data points of each galaxy in the sample (provided that these points were well outside each galaxy's optical disk) were found to obey the very compact relation
\begin{equation}
\left(\frac{v^2}{R}\right)_{\rm LAST} =\frac{N^*\beta^*c^2}{R^2}+
\frac{N^*\gamma^*c^2}{2}+\frac{\gamma_0c^2}{2}-\kappa c^2R.
\label{1.1}
\end{equation} 
The parameters  in (\ref{1.1}) were found to take the values
\begin{align}
 \beta^*=\frac{M_{\odot}G_{\rm N}}{c^2}=1.48\times10^5~{\rm cm},~~\gamma^*=5.42\times 10^{-41}~{\rm cm}^{-1},~~\gamma_0=3.06\times
10^{-30}~{\rm cm}^{-1},~~\kappa=9.54\times
10^{-54}~{\rm cm}^{-2},
\label{1.2}
\end{align} 
with $N^{*}M_{\odot}$ being the visible galactic mass in solar mass units, and $R$ being the distance of each data point from the center of the galaxy in which it is located. Other than the visible masses (viz. luminosities) of the galaxies the relation given in (\ref{1.1}) only contains four parameters and yet accounts for 138 data points. With most spiral galaxies having no more than of order $10^{10}$ to  $10^{11}$ stars, and with $\gamma_0/\gamma^*=5.6\times 10^{10}$ we see that the dominant contribution to (\ref{1.1}) is between $\gamma_0c^2/2$ and $\gamma_0c^2$, with the measured $v^2/c^2R$ at the last data points falling right in this range. The $\kappa$ term only affects the 20 or so galaxies in the sample that go out to very large $R$. With the Newtonian term falling at large distances, the $N^*\gamma^*$, $\gamma_0$ and $\kappa$ terms control the centripetal accelerations at the last data points. The universality of (\ref{1.1}) is not just in its universal parameters, it is also in the fact that the observed last data point in any given galaxy is not determined by the galaxy itself but by the limits of observations that are made at distances that differ for different galaxies. Thus (\ref{1.1}) holds no matter where the observed data points might be found to cut  off.

In (\ref{1.1}) we recognize two $N^*$-dependent $\beta^*$ and $\gamma^*$ terms that depend only on the amount of visible mass within each galaxy, while also recognizing  two other terms ($\gamma_0$ and $\kappa$) that are 
$N^*$-independent and completely universal. Thus from a knowledge solely of these four parameters and the visible mass content of any spiral galaxy one is able to determine the rotational velocities of the visible material, with no galactic  dark matter or any of its free parameters being required. With $\gamma_0$ being found to be of order the inverse of the Hubble radius and with $\kappa$ being found to be of order a typical cluster of galaxies scale ($\kappa^{-1/2}\sim 105 $ Mpc), the values for $\gamma_0$ and $\kappa$ that are obtained show that they are of order the cosmological scales that are associated with the homogeneous Hubble flow and the inhomogeneities in it. 

As presented above,  (\ref{1.1}) is a purely phenomenological relation that holds regardless of which specific theory of gravity one might wish to consider, a relation that  would thus need to be derived in any chosen theory. In the conformal gravity theory of interest to us in this paper (\ref{1.1}) has actually been derived theoretically as an exact relation in the theory, with the $\gamma_0$ and $\kappa$ terms indeed being generated by the cosmological background and the fluctuations around it, material that is indeed outside of the galaxies of interest, and thus naturally independent of $N^*$. We discuss how this comes  about in Secs. \ref{S2} and \ref{S3}. Also in  Sec. \ref{S3} we look at the implications of the length scale associated with $\kappa$ for the recombination era cosmic microwave background (CMB), and identify the temperatures (respectively $10^{9\circ}$K and $10^{13\circ}$K) at which conformal gravity metric and acoustic signals need to originate in order to fill out the recombination era sky. In Sec. \ref{S4} we study the growth of metric fluctuation amplitudes in the conformal theory, to show that they grow quite substantially. In an appendix we present a simplified discussion of our analysis.
\section{Conformal gravity}
\label{S2}

Conformal gravity has been advanced as a candidate alternate theory of gravity (see e.g. the reviews \cite{Mannheim2006,Mannheim2017} and references therein). The theory is a pure pure metric theory of gravity that possesses all of the general coordinate invariance and equivalence principle structure of standard  gravity while augmenting it with an additional symmetry, local conformal invariance, in which  the action is left invariant under local conformal transformations on the metric of the form $g_{\mu\nu}(x)\rightarrow e^{2\alpha(x)}g_{\mu\nu}(x)$ with arbitrary local phase $\alpha(x)$. Under such a symmetry a gravitational sector action that is to be a polynomial function of the Riemann tensor is uniquely prescribed, and with use of the Gauss-Bonnet theorem is given by (see e.g. \cite{Mannheim2006}) 
\begin{eqnarray}
I_{\rm W}=-\alpha_g\int d^4x\, (-g)^{1/2}C_{\lambda\mu\nu\kappa}
C^{\lambda\mu\nu\kappa}
\equiv -2\alpha_g\int d^4x\, (-g)^{1/2}\left[R_{\mu\kappa}R^{\mu\kappa}-\frac{1}{3} (R^{\alpha}_{\phantom{\alpha}\alpha})^2\right].
\label{2.1}
\end{eqnarray}
Here $\alpha_g$ is a dimensionless  gravitational coupling constant, and
\begin{eqnarray}
C_{\lambda\mu\nu\kappa}= R_{\lambda\mu\nu\kappa}
-\frac{1}{2}\left(g_{\lambda\nu}R_{\mu\kappa}-
g_{\lambda\kappa}R_{\mu\nu}-
g_{\mu\nu}R_{\lambda\kappa}+
g_{\mu\kappa}R_{\lambda\nu}\right)
+\frac{1}{6}R^{\alpha}_{\phantom{\alpha}\alpha}\left(
g_{\lambda\nu}g_{\mu\kappa}-
g_{\lambda\kappa}g_{\mu\nu}\right)
\label{2.2}
\end{eqnarray}
is the conformal Weyl tensor. (Here and throughout we follow the notation and conventions of \cite{Weinberg1972} in which the signature of the metric is $(-1,1,1,1)$.) Functional variation of $I_{\rm W}$ with respect to the metric generates fourth-order derivative gravitational equations of motion for the metric of the form (see e.g. \cite{Mannheim2006}) 
\begin{eqnarray}
-\frac{2}{(-g)^{1/2}}\frac{\delta I_{\rm W}}{\delta g_{\mu\nu}}=4\alpha_g W^{\mu\nu}=4\alpha_g\left[2\nabla_{\kappa}\nabla_{\lambda}C^{\mu\lambda\nu\kappa}-
R_{\kappa\lambda}C^{\mu\lambda\nu\kappa}\right]=4\alpha_g\left[W^{\mu
\nu}_{(2)}-\frac{1}{3}W^{\mu\nu}_{(1)}\right]=T^{\mu\nu},
\label{2.3}
\end{eqnarray}
where the functions $W^{\mu \nu}_{(1)}$ and $W^{\mu \nu}_{(2)}$ (respectively associated with the $(R^{\alpha}_{\phantom{\alpha}\alpha})^2$ and $R_{\mu\kappa}R^{\mu\kappa}$ terms in (\ref{2.1})) are given by
\begin{eqnarray}
W^{\mu \nu}_{(1)}&=&
2g^{\mu\nu}\nabla_{\beta}\nabla^{\beta}R^{\alpha}_{\phantom{\alpha}\alpha}                                             
-2\nabla^{\nu}\nabla^{\mu}R^{\alpha}_{\phantom{\alpha}\alpha}                          
-2 R^{\alpha}_{\phantom{\alpha}\alpha}R^{\mu\nu}                              
+\frac{1}{2}g^{\mu\nu}(R^{\alpha}_{\phantom{\alpha}\alpha})^2,
\nonumber\\
W^{\mu \nu}_{(2)}&=&
\frac{1}{2}g^{\mu\nu}\nabla_{\beta}\nabla^{\beta}R^{\alpha}_{\phantom{\alpha}\alpha}
+\nabla_{\beta}\nabla^{\beta}R^{\mu\nu}                    
 -\nabla_{\beta}\nabla^{\nu}R^{\mu\beta}                       
-\nabla_{\beta}\nabla^{\mu}R^{\nu \beta}                          
 - 2R^{\mu\beta}R^{\nu}_{\phantom{\nu}\beta}                                    
+\frac{1}{2}g^{\mu\nu}R_{\alpha\beta}R^{\alpha\beta},
\label{2.4}
\end{eqnarray}                                 
and where $T^{\mu\nu}$ is the conformal invariant and thus traceless energy-momentum tensor associated with a conformal matter source. (The form $W^{\mu\nu}=2\nabla_{\kappa}\nabla_{\lambda}C^{\mu\lambda\nu\kappa}-R_{\kappa\lambda}C^{\mu\lambda\nu\kappa}$ was given by Bach \cite{Bach1921}, while the form for $W^{\mu \nu}_{(1)}$ and $W^{\mu \nu}_{(2)}$ may be found in \cite{DeWitt1964}.) With (\ref{2.3}) being a fourth-order derivative equation for the metric, familiarity with higher-derivative theories raises the concern that  the theory might have instability or unitarity problems. However, it turns out \cite{Bender2008a,Bender2008b} that the Hamiltonian of the theory is not Hermitian but is instead $PT$ symmetric, with use  of the $PT$ norm, viz. the overlap of a ket with its $PT$ conjugate rather than with its Hermitian conjugate, then leading to a quantum conformal gravity theory with neither negative energy or negative norm states \cite{Bender2008a,Bender2008b}. In addition, with the  coupling constant $\alpha_g$ being dimensionless, as a quantum theory conformal gravity is renormalizable \cite{Stelle1977,Fradkin1985}. Thus just like the way that classical electrodynamics can be derived from quantum electrodynamics, the classical conformal gravity theory that we study here can be associated with matrix elements of quantum gravitational operators as evaluated in states with an indefinite number of gravitons that couple with a renormalized gravitational coupling constant. 

The utility of having a theory that is based on the Weyl tensor is that its properties  enable us to distinguish between the background cosmology and the fluctuations around it. Specifically, the Weyl tensor vanishes in geometries that are conformal to flat, so that in the homogeneous and isotropic conformal to flat background Robertson-Walker and de Sitter geometries of interest to cosmology the background $W^{\mu\nu}$ is zero. Despite this,  fluctuations (i.e., cosmological inhomogeneities such as the galaxies and stars of interest to us in this paper) around that background are not conformal to flat, with the fluctuating $ W^{\mu\nu}$ then not being zero. 

For a static, spherically symmetric source both the Weyl tensor  and $W^{\mu\nu}$  do not vanish, and thus we have to solve the fourth-order differential equation $W^{\mu\nu}=T^{\mu\nu}/4\alpha_g$.  As is standard, using general coordinate invariance we can reduce the metric in a static, spherically symmetric geometry to just two independent metric coefficients. Then, as noted in \cite{Mannheim1989}, through use of the underlying conformal symmetry that the theory possesses we can reduce the metric even further to just one independent coefficient (designated $B(r)$), to give a line element of the form
\begin{eqnarray}
ds^2 = B(r)c^2dt^2 -\frac{dr^2}{B(r)} -r^2d\theta^2-r^2\sin^2\theta d\phi^2.
\label{2.5}
\end{eqnarray}
Given (\ref{2.5}) we can determine the components of $W_{\mu\nu}$ in a closed form, to then yield \cite{Mannheim1994} the exact and very compact relation 
\begin{eqnarray}                                                                               
\frac{3}{B}\left(W^0_{{\phantom 0} 0}-W^r_{{\phantom r} r}\right)=B^{\prime\prime\prime\prime}+\frac{4}{r}B^{\prime\prime\prime}=\frac{3}{4\alpha_g B}\left(T^0_{{\phantom 0} 0}
-T^r_{{\phantom r} r}\right) =f(r),
\label{2.6}
\end{eqnarray}      
with (\ref{2.6}) serving to define $f(r)$.
The solution to (\ref{2.6}) can readily be obtained in closed form and is given by  \cite{Mannheim1994} 
\begin{eqnarray}
B(r)&=&B_0(r)-\frac{r}{2}\int_0^r
dr^{\prime}r^{\prime 2}f(r^{\prime})
-\frac{1}{6r}\int_0^r
dr^{\prime}r^{\prime 4}f(r^{\prime})
-\frac{1}{2}\int_r^{\infty}
dr^{\prime}r^{\prime 3}f(r^{\prime})
-\frac{r^2}{6}\int_r^{\infty}
dr^{\prime}r^{\prime }f(r^{\prime}).
\label{2.7}
\end{eqnarray}                                 
As we see, $B(r)$ receives contributions from matter distributions that can extend all the way to $r=\infty$, matter distributions that play no role in the Newtonian case.  In (\ref{2.7}) we have also included a  $B_0(r)$ contribution that identically satisfies $B_0^{\prime\prime\prime\prime}+(4/r)B_0^{\prime\prime\prime}=0$ at all points $r$, even including those points where $f(r)$ is nonzero. 

The $f(r)$-dependent term contains two types of contributions, those due to the $\int_0^r$ terms and those due to the $\int_r^{\infty}$ terms, terms that are respectively associated with the local material within galaxies and with the global material exterior to them. While below we will take the spiral (i.e., nonspherical) nature of galaxies composed of $N^*$ stars into consideration,  for the moment we shall take the nearby matter to consist of a single static, spherically symmetric star of radius $R_0$ with its center situated at the origin of the coordinate system that we are using,  so that its contribution to $f(r)$ is only nonzero in an $r<R_0$ region. Similarly, we shall take the global contribution to begin at some cluster of galaxies scale $R_1$. There is thus a region $R_0<r<R_1$ in which $f(r)$ is zero. In this region the $f(r)$-dependent contribution to $B(r)$, which we label $B^*(r)$,  takes the form
\begin{eqnarray}
B^*(R_0<r<R_1)&=&-\frac{2\beta^*}{r}+\gamma^*r+w^*-\kappa^*r^2,
\label{2.8}
\end{eqnarray}                                 
where
\begin{align}
-2\beta^*&= -\frac{1}{6}\int_0^{R_0}
dr^{\prime}r^{\prime 4}f(r^{\prime}),\quad
\gamma^*=-\frac{1}{2}\int_0^{R_0}
dr^{\prime}r^{\prime 2}f(r^{\prime}),\quad
w^*=-\frac{1}{2}\int_{R_1}^{\infty}dr^{\prime}r^{\prime 3}f(r^{\prime}),\quad
\kappa^*=\frac{1}{6}\int_{R_1}^{\infty}dr^{\prime}r^{\prime }f(r^{\prime}).
\label{2.9}
\end{align}                                 
(The matter-free region solution given in (\ref{2.8}) may also be found in \cite{Riegert1984b}, with the connection to the matter source $f(r)$ as given in (\ref{2.9}) being found in \cite{Mannheim1994}.)

With $B^*(r)$ reducing to $B^*(r)=w^*-2\beta^*/r$ when $r$ is small we thus recover the Schwarzschild metric for small $r$. Thus we can consider the solution given in (\ref{2.8}) as being a generalization of the Schwarzschild metric. Thus as required by covariance we actually generalize the relativistic Schwarzschild solution rather than the nonrelativistic Newton's Law of Gravity. Since we do obtain the Schwarzschild solution for small $r$ the conformal gravity theory meets the classic solar system tests of general relativity, doing so  even as it does not contain the Einstein equations. To be in agreement with solar system data a theory only needs to be able to recover the solutions to the Einstein equations on solar system distance scales without needing to recover the Einstein equations themselves. However, solar system data do not require that the solution not receive additional, no longer negligible,  contributions at distances beyond the solar system. In the conformal theory there are such additional contributions, with the $\gamma^*r-\kappa^*r^2$ term only being important at large $r$. As we now discuss, its presence is central to the ability of the conformal theory to explain the systematics of galactic rotation curves without the need for any galactic dark matter.

The two local terms, viz. those that depend on the matter distribution in $r<R_0$,  give a stellar potential of the form
\begin{equation}
V^{*}(r)=-\frac{\beta^{*}c^2}{r}+\frac{\gamma^{*}c^2 r}{2}.
\label{2.10}
\end{equation}
Thus for a spiral galaxy with $N^*$ stars, surface brightness $\Sigma (R)=\Sigma_0e^{-R/R_0}$, total luminosity $L=2\pi \Sigma_0 R_0^2$,  mass to light ratio $M/L$ and mass $N^*M_{\odot}=(M/L)2\pi \Sigma_0R_0^2$, we obtain a local, weak gravity, galactic centripetal acceleration of the form  \cite{Mannheim2006} 
\begin{eqnarray}
\frac{v_{{\rm LOC}}^2}{R}&=&
\frac{N^*\beta^*c^2 R}{2R_0^3}\left[I_0\left(\frac{R}{2R_0}
\right)K_0\left(\frac{R}{2R_0}\right)-
I_1\left(\frac{R}{2R_0}\right)
K_1\left(\frac{R}{2R_0}\right)\right]
\nonumber \\
&&+\frac{N^*\gamma^* c^2R}{2R_0}I_1\left(\frac{R}{2R_0}\right)
K_1\left(\frac{R}{2R_0}\right),
\label{2.11}
\end{eqnarray} 
due to this potential, with asymptotic limit
\begin{eqnarray}
\frac{v_{{\rm LOC}}^2}{R} \rightarrow \frac{N^*\beta^*c^2}{R^2}+
\frac{N^*\gamma^*c^2}{2},
\label{2.12}
\end{eqnarray} 
just as required of (\ref{1.1}).

Similarly, the two global terms in (\ref{2.8}) give a weak gravity, galactic centripetal acceleration of the form
\begin{eqnarray}
\frac{v_{{\rm GLOB}}^2}{R} =-\kappa^*c^2R,
\label{2.13}
\end{eqnarray} 
again just as required of (\ref{1.1}). Thus  in  (\ref{1.1}) we can identify $\kappa=(1/6) \int_{R_1}^{\infty}dr^{\prime}r^{\prime }f(r^{\prime})$ as being due to large scale inhomogeneities, with its measured $105$ Mpc value indeed being associated with an inhomogeneous cosmological scale. Thus in the conformal gravity theory we can use galactic rotation curve data to measure the inhomogeneity scale of the universe.

In regard to the $B_0(r)$ term, we note that the $f(r)$-dependent term is due to sources that are associated with an expressly nonvanishing $W_{\mu\nu}$ (i.e., inhomogeneities in the cosmological background), while the $B_0(r)$ term is associated with sources for which $W_{\mu\nu}$ expressly vanishes everywhere (just as required of the cosmological background itself). There are two ways in which the fourth-order derivative function $W_{\mu\nu}$ could vanish identically everywhere: the Weyl tensor itself could vanish or the Weyl tensor could  obey $2\nabla_{\kappa}\nabla_{\lambda}C^{\mu\lambda\nu\kappa}-R_{\kappa\lambda}C^{\mu\lambda\nu\kappa}=0$. With all the nonvanishing components of  the Weyl tensor being proportional to $ B^{\prime\prime}-2B^{\prime}/r+2(B-1)/r^2$ in the metric given in (\ref{2.5}), for $C^{\mu\lambda\nu\kappa}=0$ the solution is given as 
\begin{eqnarray}
B_0(r)=1+\gamma_0r-K_0r^2.
\label{2.14}
\end{eqnarray}
For $2\nabla_{\kappa}\nabla_{\lambda}C^{\mu\lambda\nu\kappa}-R_{\kappa\lambda}C^{\mu\lambda\nu\kappa}=0$ the two other solutions required  for the fourth-order derivative $W_{\mu\nu}=0$ equation are given by $B_0(r)=w_0+u_0/r$.  Since this solution has to hold everywhere a finiteness condition at $r=0$ excludes the $u_0/r$ term. On incorporating the $K_0r^2$ term into the $\kappa^* r^2$ term given in (\ref{2.8}) we thus obtain
 \begin{eqnarray}
 B_0(r)=1+w_0+\gamma_0r 
 \label{2.15}
 \end{eqnarray}
 as the relevant $B_0(r)$ term.
 
To see in which way (\ref{2.15}) is in fact cosmological in nature, we note that 
when a comoving time Robertson-Walker geometry is written in spatially isotropic coordinates the line element is given by  
\begin{eqnarray}
ds^2=c^2dt^2-\frac{a^2(t)}{(1+k\rho^2/4)^2}[d\rho^2+\rho^2d\theta^2+\rho^2\sin^2\theta d\phi^2],
\label{2.16} 
\end{eqnarray}
where $a(t)$ is the expansion radius and $k$ is the spatial three-curvature. If we now make the coordinate transformation  \cite{Mannheim1989}
\begin{eqnarray}
\rho&=&\frac{8+4\gamma_0 r-8(1+\gamma_0r)^{1/2}}{\gamma_0^2r},\quad
\tau=\int \frac{dt}{a(t)},
\label{2.17}
\end{eqnarray}
 (\ref{2.16}) then takes the coordinate-equivalent form
\begin{eqnarray}
ds^2=a^2(\tau)\bigg{[}c^2d\tau^2-\frac{dr^2}{(1+\gamma_0r)^2}-\frac{r^2}{(1+\gamma_0r)}(d\theta^2+\sin^2\theta d\phi^2)\bigg{]},
\label{2.18} 
\end{eqnarray}
where $\gamma_0^2=-4k$. With $\gamma_0$ being real and nonzero, it follows that the only allowed value of $k$ for which we can in fact make this transformation at all is negative.
Noting now that the Weyl tensor not only vanishes in a Robertson-Walker geometry but also in a conformally transformed one, we now conformally transform (\ref{2.18}) by multiplying the line element by $\Omega^2(r,\tau)$ where $\Omega(r,\tau)=(1+\gamma_0r)^{1/2}/a(\tau)$, with (\ref{2.18}) then taking the form
\begin{eqnarray}
ds^2=(1+\gamma_0r)c^2d\tau^2-\frac{dr^2}{(1+\gamma_0r)}-r^2(d\theta^2+\sin^2\theta d\phi^2).
\label{2.19} 
\end{eqnarray}
This is precisely a static, spherically symmetric line element with $B_0(r)=1+\gamma_0r$, i.e., just as given by (\ref{2.15}) above. Thus when written in galactic rest frame coordinates a comoving Robertson-Walker cosmology looks just like the universal linear potential given in (\ref{1.1}) provided the spatial three-curvature of the Robertson-Walker geometry is negative. Thus the parameter $\gamma_0$ given in (\ref{1.1}) should be of cosmological scale, just as is indeed  found phenomenologically when (\ref{1.1}) is applied to galactic rotation curve data. We should note that while numerically $\gamma_0$ is found to be of order the inverse of the Hubble radius, this is only a heuristic connection,  since physically it is impossible for a $\gamma_0$ that is constant to be equal to a Hubble radius that is epoch dependent. However, the spatial three-curvature $k$ of the universe is not epoch dependent, and thus its identification with the constant $\gamma_0$ is physically legitimate.

Evidence that the spatial three-curvature $k$ of the universe is indeed negative in the conformal gravity theory has been provided by a study given in \cite{Mannheim2006} of the accelerating universe data \cite{Riess1998,Perlmutter1999} and is described below. In this study a non-fine-tuned fit to the data was presented of quality comparable to that of the fit using the fine-tuned 30\% dark matter 70\% dark energy standard concordance model established in \cite{Bahcall1999,deBernadis2000,Tegmark2004}. The reason that the conformal theory needs no fine tuning is because the underlying conformal symmetry controls the cosmological constant, making it zero at the level of the Lagrangian and making it no bigger or smaller than any other term in $T_{\mu\nu}$ after masses are generated  spontaneously (the only way to generate masses in a conformal theory) since the traceless $T_{\mu\nu}$ remains traceless (see e.g. \cite{Mannheim2017}).
Finally,  since $\gamma_0^2=-4k$, a measurement of $\gamma_0$ is a measurement of $k$. Thus in the conformal gravity theory we can use galactic rotation curve data to measure the spatial curvature of the universe.

Putting everything together, for a spiral galaxy we have a net centripetal acceleration of the form 
\begin{eqnarray}
\frac{v^2}{R}&=&
\frac{N^*\beta^*c^2 R}{2R_0^3}\left[I_0\left(\frac{R}{2R_0}
\right)K_0\left(\frac{R}{2R_0}\right)-
I_1\left(\frac{R}{2R_0}\right)
K_1\left(\frac{R}{2R_0}\right)\right]
\nonumber \\
&&+\frac{N^*\gamma^* c^2R}{2R_0}I_1\left(\frac{R}{2R_0}\right)
K_1\left(\frac{R}{2R_0}\right)+\frac{\gamma_0c^2}{2}-\kappa c^2R,
\label{2.20}
\end{eqnarray} 
 a form whose large $R$ limit is precisely (\ref{1.1}).
With just the same four parameters as given in (\ref{1.2}), in \cite{Mannheim2011,Mannheim2012a,O'Brien2012} (\ref{2.20}) has been used to fit not just the last data point in each galaxy but in fact  the entire set of all of the data points in  the 138 galaxy sample (of the order of 3000 or so data points in total), with no need for any dark matter or any of the 276 fitting parameters that  it needs for the 138 galaxy sample. Thus the $N^*\gamma^*$, $\gamma_0$ and $\kappa$ terms replace dark matter.

We note that in any theory in which the gravitational potential is no longer just a $1/r$ potential, Newton's decoupling theorem will no longer hold (for a $1/r$ potential the force falls like $1/r^2$ leading to exterior region cancellations since the solid angle grows like $r^2$, with there being no such cancelations for any other potential  since the solid angle does not change even as the force does). Thus any modification of Newton's Law of Gravity (essentially the only alternative to dark matter) will lead to some form of external gravitational contribution to interior region galactic dynamics. As respectively described in \cite{Chae2020} and \cite{Mannheim2021} such effects have been detected in Milgrom's MOND theory \cite{Milgrom1983a} and in Moffat's MOG theory \cite{Moffat2006}. However, these effects are found to be quite small and not competitive with the contributions of the conformal gravity $N^*\gamma^*$, $\gamma_0$ and $\kappa$ terms that are associated with potentials that are big on large distance scales. 

Since the linear potentials put out by individual sources do grow with distance (cf. $V^{*}(r)=-\beta^{*}c^2/r+\gamma^{*}c^2 r/2$), the potentials due to the material most distant from a given galaxy will also grow with distance, and thus  have an impact within any given galaxy, just as we see with the $\gamma_0$ and $\kappa$ terms. These terms are naturally not just independent of $N^*$ (i.e. they have nothing to do with the mass content of any given galaxy),  but are also naturally cosmological in scale, since the most distant sources provide the largest linear potential terms. Hence in conformal gravity there is a natural interplay between local and global physics. This effect is absent in the standard dark matter model, and introducing dark matter within galaxies can be understood purely as an attempt to describe global physics in local terms. In the conformal gravity theory local dark matter is replaced by the contribution of the rest of the universe. Thus the galactic missing mass is not missing at all, it is the rest of the visible universe and it has been hiding in plain sight all along.

Finally, we see that when a test particle travels on a rotational orbit in a galaxy it is sampling not just the local gravitational field  ($\beta^*$ and $\gamma^*$) but also the global gravitational field ($\gamma_0$ and $\kappa$) and allows us to measure it. And with the value of $\kappa$ now having been determined we now show how this scale is manifest in the anisotropy in the CMB in the recombination era.

\section{$\kappa$ and the recombination era  cosmic microwave background}
\label{S3}

In studying recombination era anisotropies in the CMB there are two proper distances that play a central role. The first is the proper angular diameter distance of a candidate yardstick between endpoints that are at the same recombination distance $r_R$ and recombination time $t_R$ as seen by an observer at $r_0=0$ at time $t_0$, i.e., at coordinates that satisfy
\begin{eqnarray}
\int_0^{r_R} \frac{dr}{(1-kr^2)^{1/2}}=\int_{t_R}^{t_0}\frac{cdt}{a(t)}.
\label{3.1}
\end{eqnarray}
With the endpoints of the yardstick being separated by an angle $\theta$ as subtended at the observer, the proper angular diameter of the yardstick is given by \cite{Weinberg1972}
\begin{eqnarray}
d(\theta)=2a(t_R)\int_0^{r_R\sin(\theta/2)} \frac{dr}{(1-kr^2)^{1/2}}.
\label{3.2}
\end{eqnarray}
The other proper distance is the size to which a fluctuation produced at time $t_F$ grows at recombination. There are two  distances of interest,  a metric one (velocity $c$), and an acoustic one (approximate velocity $c_s=c/\sqrt{3}$), viz.
\begin{eqnarray}
D(t_F,t_R)=a(t_R)\int_{t_F}^{t_R}\frac{cdt}{a(t)},\qquad D(t_F,t_R,c_S)=\frac{c_S}{c}D(t_F,t_R).
\label{3.3}
\end{eqnarray}
For the universe to be horizon free $D(0,t_R)$ would have to be at least as big as $d(\pi)$.

For the standard radiation-dominated Robertson-Walker cosmology with $k=0$ and $a(t)=At^{1/2}$ a fluctuation that starts out at $t_F=0$ will grow to a proper distance $D(0,t_R)= 2ct_R$ at recombination. To an observer at $t_0$ who receives signals propagating from recombination in a standard matter-dominated Robertson-Walker cosmology with $k=0$ and $a(t)=Bt^{2/3}$ where $B=At_R^{-1/6}$ will measure a proper angular diameter given by $d(\pi)=6ct_R^{2/3}t_0^{1/3}$. (We have set $t_0\gg t_R$, and for simplicity we have taken the radiation domination era to transit to matter domination at recombination.) Assuming that $a(t)T(t)$ is a time-independent constant, we obtain $D(0,t_R)/d(\pi)=(T_0/T_R)^{1/3}/3$, where $T_R$ is the recombination temperature and $T_0\sim 3^{\circ}$K is the current temperature. With $T_0/T_R=10^{-3}$ we see that $D(0,t_R)/d(\pi)$ is much less than one. With a matching of  $D(0,t_R)$ and $d(\theta)$ only occurring at a few degrees or so, this then is the standard model horizon problem. To resolve it within standard gravity one would need a period of much more rapid expansion prior to $t_R$, and this can be supplied by a period of exponential expansion \cite{Brout1978,Starobinsky1979,Kazanas1980,Guth1981}  in the very early universe \cite{footnote1}. For the conformal gravity theory the expansion rate of $a(t)$ is far different from either $At^{1/2}$ or $Bt^{2/3}$, and, as we now show, this enables conformal cosmology to be horizon free, without the need for any period of exponential expansion in the very early universe.

To determine numerical conformal gravity predictions we need to determine some background cosmological parameters. To this end we note that since an unbroken conformal symmetry entails no mass or length scales, such scales have to arise via spontaneous breakdown. While this is done dynamically via the nonvanishing of $\langle \Omega|\bar{\psi}\psi |\Omega\rangle$ (see \cite{Mannheim2017}) we can simulate it by using a scalar field $S(x)$ that acquires a constant value $S_0$. Using fermions to be representative of the matter fields, 
for the fermion and scalar field sector we take the curved space matter action to be of the form 
\begin{eqnarray}
I_M=-\int d^4x(-g)^{1/2}\left[\frac{1}{2}\nabla_{\mu}S\nabla^{\mu}S-\frac{1}{12}S^2R^\mu_{\phantom{\mu}\mu}
+\lambda S^4+
i\bar{\psi}\gamma^{c}V^{\mu}_c(x)[\partial_\mu+\Gamma_\mu(x)]\psi-hS\bar{\psi}\psi\right],
\label{3.4}
\end{eqnarray}                                 
where $h$ and $\lambda$ are dimensionless coupling constants.  Here $V^{\mu}_c(x)$ is a vierbein and $\Gamma_{\mu}(x)=-(1/8)[\gamma_a,\gamma_b](V^b_{\nu}\partial_{\mu}V^{a\nu}+V^b_{\lambda}\Gamma^{\lambda}_{\phantom{\lambda}\nu\mu}V^{a\nu})$ is the spin connection, with the $\gamma_a$ being a set of fixed axis Dirac gamma matrices and $\Gamma^{\lambda}_{\phantom{\lambda}\nu\mu}$ being $(1/2)g^{\lambda\sigma}[\partial_{\mu}g_{\sigma\nu} +\partial_{\nu}g_{\sigma\mu} -\partial_{\sigma}g_{\mu\nu} ]$. As such, this action is locally conformal invariant under $g_{\mu\nu}(x)\rightarrow e^{2\alpha(x)}g_{\mu\nu}(x)$, $V^a_{\mu}(x)\rightarrow e^{\alpha(x)}V^a_{\mu}(x)$, $\psi(x)\rightarrow e^{-3\alpha(x)/2}\psi(x)$, $S(x)\rightarrow e^{-\alpha(x)}S(x)$.  Thus the very conformal symmetry that makes $I_{\rm W}$ be fourth order also makes $I_M$ be second order, just as $I_{M}$  is taken to be in the standard cosmological theory. Variation of
the $I_M$  action with respect to  $\psi(x)$ and $S(x)$ yields the equations of motion
\begin{eqnarray}
i\gamma^{c}V^{\mu}_c(x)[\partial_\mu+\Gamma_\mu(x)]\psi - h S \psi = 0,
\label{3.5}
\end{eqnarray}                                 
\begin{eqnarray}
\nabla_{\mu}\nabla^{\mu}S+\frac{1}{6}SR^\mu_{\phantom{\mu}\mu}
-4\lambda S^3 +h\bar{\psi}\psi=0,
\label{3.6}
\end{eqnarray}                                 
while variation with respect to the metric yields an energy-momentum tensor that takes the form \cite{Mannheim2006}
\begin{equation}
T^{\mu \nu} = \frac{1}{c}\left[(\rho_m+p_m)U^{\mu}U^{\nu}+p_mg^{\mu\nu}\right] 
-\frac{1}{6}S_0^2\left(R^{\mu\nu}
-\frac{1}{2}g^{\mu\nu}R^\alpha_{\phantom{\alpha}\alpha}\right)         
-g^{\mu\nu}\lambda S_0^4,
\label{3.7}
\end{equation}                                 
when $S$ is equal to a constant $S_0$ and the fermion sector forms a perfect fluid with fluid four-velocity $U_{\mu}$, fluid energy density $\rho_m$ and fluid pressure $p_m$. With $W^{\mu\nu}$ vanishing in a Robertson-Walker geometry,  according to (\ref{2.3}) the cosmological $T^{\mu\nu}$ would have to obey $T^{\mu\nu}=0$, to thus yield
\begin{equation}
\frac{1}{6}S_0^2\left(R^{\mu\nu}
-\frac{1}{2}g^{\mu\nu}R^\alpha_{\phantom{\alpha}\alpha}\right) = 
\frac{1}{c}\left[(\rho_m+p_m)U^{\mu}U^{\nu}+p_mg^{\mu\nu}\right]  -g^{\mu\nu}\lambda S_0^4.
\label{3.8}
\end{equation}                                 
 We recognize the conformal cosmological evolution equation given in  (\ref{3.8}) as being of the form
of none other than the cosmological evolution equation of the standard theory, save only for the fact that for cosmology the standard attractive Newtonian $G_{{\rm N}}$ has been replaced by an effective, dynamically induced one given by
\begin{equation}
G_{{\rm eff}}=-\frac{3c^3}{4\pi S_0^2},
\label{3.9}
\end{equation}                                 
viz.  by an effective cosmological gravitational coupling constant that, as had been noted in \cite{Mannheim1992}, is expressly negative. With this negative coupling constant conformal cosmology naturally has a  repulsive component, which is why it can fit the accelerating universe data without fine tuning. With the local Newtonian $G_{{\rm N}}$ in the conformal gravity theory being associated with the fourth-moment integral of the source $f(r)$  as given in (\ref{2.9}) (viz. the $\beta^*=M_{\odot}G_{\rm N}/c^2$ coefficient), in the conformal theory the local (Weyl tensor nonzero) and global (Weyl tensor zero) gravitational coupling constants are independent, to thus release cosmology from being controlled by the local Newtonian $G_{{\rm N}}$, to thereby provide options for cosmology that cannot be obtained if global cosmology is controlled by the local Newtonian $G_{{\rm N}}$.

To be able to see how central the negative sign of $G_{{\rm eff}}$ is to cosmic acceleration 
we define 
\begin{equation}
\bar{\Omega}_{M}(t)=\frac{8\pi G_{{\rm eff}}\rho_{m}(t)}{3c^2H^2(t)}, \quad
\bar{\Omega}_{\Lambda}(t)=\frac{8\pi G_{{\rm
eff}}\Lambda}{3cH^2(t)}=-\frac{2c^2\Lambda}{S_0^2H^2(t)},\quad \bar{\Omega}_k(t)=-\frac{kc^2}{\dot{a}^2(t)},
\label{3.10}
\end{equation}                                 
where $H=\dot{a}/a$ and $\Lambda=\lambda S_0^4$. And on introducing the deceleration parameter $q=-a\ddot{a}/\dot{a}^2$, from (\ref{3.8}) and (\ref{3.10}) we obtain
\begin{eqnarray}
\dot{a}^2(t) +kc^2
&=&\dot{a}^2(t)\left(\bar{\Omega}_{M}(t)+
\bar{\Omega}_{\Lambda}(t)\right),\quad \bar{\Omega}_M(t)+
\bar{\Omega}_{\Lambda}(t)+\bar{\Omega}_k(t)=1,
\nonumber \\
q(t)&=&\frac{1}{2}\left(1+\frac{3p_m}{\rho_m}\right)\bar{\Omega}_M(t)
-\bar{\Omega}_{\Lambda}(t)
\label{3.11}
\end{eqnarray}
as the  evolution equations of conformal cosmology. 

In the conformal theory we must take $\lambda$ and $\Lambda$ to be negative because there is a release of free energy at the spontaneous breakdown phase transition (at temperature $T_V=10^{15\circ}$K or so), with the energy at later times, i.e., at lower temperatures,  being lower not higher. If we set  $\Lambda=-aT_V^4/c$ where $a$ is the black-body constant we obtain  $\Lambda=-2.52\times 10^{35}$ gm sec$^{-1}$ cm$^{-2}$.  While $\Lambda$ is negative we recall that so is $G_{\rm eff}$. Consequently, $\bar{\Omega}_{\Lambda}(t)$ is positive. With such a large value for $\Lambda$, the perfect fluid $\bar{\Omega}_{M}(t)$ contribution to cosmic evolution is suppressed with respect to the $\bar{\Omega}_{\Lambda}(t)$ contribution other than at temperatures as high as $10^{15\circ}$K. The matter field $\bar{\Omega}_{M}(t)$ thus plays no role at the $T_R$ temperature of interest to recombination, and it is thus immaterial as to whether we model it by a relativistic or a nonrelativistic fluid except at temperatures of the order $10^{15\circ}$K,  where the fluid would be relativistic. Then, with $k$ being negative,  in the $T\ll T_V$  region where $\bar{\Omega}_M(t)$ is negligible it follows from (\ref{3.11}) that $\bar{\Omega}_{\Lambda}(t)$  is constrained to have to lie in the narrow range $0<\bar{\Omega}_{\Lambda}(t)<1$ no matter how big $\Lambda$ itself might be. I.e., with $k<0$ $\bar{\Omega}_{\Lambda}(t)$ has to approach one from below and not from above. That this surprising result (surprising from the perspective of standard Einstein gravity that is) is possible is because with $G_{{\rm N}}$ being replaced by a much smaller $G_{\rm eff}$ ($S_0$ being very big) the cosmological constant  does not gravitate as much as it does in the standard theory. From a measurement of the current era deceleration parameter we now show that $\bar{\Omega}_{\Lambda}(t_0)$ really is quenched sufficiently to be less than one.

With $\rho_m=3p_m=A/a^4(t)=aT^4(t)/c$ (where $A$ is a constant), and with $k$ and $\lambda$ being  negative the exact solution for the expansion radius $a(t)$ that follows from (\ref{3.11}) is given as \cite{Mannheim2006}
\begin{eqnarray}
a^2(t)= -\frac{k(\beta-1)}{2\sigma}
-\frac{k\beta{\rm sinh}^2 (\sigma^{1/2} ct)}{\sigma},
\label{3.12}
\end{eqnarray}
where the parameters $\sigma$ and $\beta$ are defined as 
\begin{eqnarray}
 \sigma =-2\lambda S_0^2=\frac{8\pi G_{\rm eff}\Lambda}{3c^3},\quad
\beta=\left(1- \frac{16A\lambda}{k^2c}\right)^{1/2}.
\label{3.13}
\end{eqnarray}
If we were to set $A=0$ then $\beta$ would be exactly equal to one. Since the radiation fluid does only contribute nontrivially  in the early universe, for nonearly universe era purposes we can set $\beta=1$, and thus obtain 
\begin{equation}
a(t)=
\left(\frac{-k}{\sigma}\right)^{1/2}\sinh(\sigma^{1/2} ct).
\label{3.14}
\end{equation}
With this form for $a(t)$ (\ref{3.10}) and (\ref{3.11}) yield  
\begin{equation}
 \bar{\Omega}_{\Lambda}(t ) = {\rm tanh}^2 (\sigma^{1/2}ct),\quad  \bar{\Omega}_k(t)= {\rm
sech}^2 (\sigma^{1/2}ct), \quad q(t) = -\bar{\Omega}_{\Lambda}(t )=-\tanh^2 (\sigma^{1/2}ct),
\label{3.15}
\end{equation}
and as we see $\bar{\Omega}_{\Lambda}(t )$ does indeed lie between zero and one, with the automatically accelerating $q(t)$ automatically lying between zero and minus one.
With the radiation fluid only contributing in the early universe and with $\bar{\Omega}_k(t)$ decreasing with $t$ while $\bar{\Omega}_{\Lambda}(t)$ increases with $t$, the cosmology thus goes through three epochs, first matter, then spatial curvature and finally cosmological constant domination, with, as we show below,  spatial curvature being found to dominate at recombination.

To show that this picture is supported by observation, given (\ref{3.14}) the relation between luminosity distance $d_L$ and redshift $z$ is found to be of the form   \cite{Mannheim2006}
\begin{equation} 
d_L=-\frac{c}{H_0}\frac{(1+z)^2}{q_0}\left[1-\left(1+q_0-
\frac{q_0}{(1+z)^2}\right)^{1/2}\right],
\label{3.16}
\end{equation}
where $q_0=q(t_0)$ is the current era value of the deceleration parameter and $H_0=H(t_0)$ is the current era value of the Hubble parameter. Fitting the accelerating universe data with (\ref{3.16})  is found to give fully acceptable fitting, with $q_0$ being fitted to a value  $-0.37$ \cite{Mannheim2006}. From (\ref{3.15}) it then follows that $\bar{\Omega}_{\Lambda}(t_0)=+0.37$, viz. right in the $0<\bar{\Omega}_{\Lambda}(t)<1$ range just as required. Given that $\Lambda=-2.52\times 10^{35}$ gm sec$^{-1}$ cm$^{-2}$, we find that with a current era value for the Hubble parameter of $H_0$=72 km sec$^{-1}$ Mpc$^{-1}$ $=2.33\times 10^{-18}$ sec$^{-1}$, $\bar{\Omega}_{\Lambda}(t_0 )=-2c^2\Lambda/S_0^2H_0^2$ takes the value $0.37$ if $S_0=1.50\times10^{46}$ gm$^{1/2}$ sec$^{-1/2}$. With this large value for $S_0$ we find that $G_{\rm eff}=-2.85\times 10^{-62}$ cm$^3$ gm$^{-1}$ sec$^{-1}$, viz. 54 orders of magnitude smaller than $G_{\rm N}=6.67\times 10^{-8}$ cm$^3$ gm$^{-1}$ sec$^{-1}$. (We thus take $c\Lambda/\rho_m\sim T_V^4/T^4$ to be very large when $T \ll T_V$, with the very  small $G_{\rm eff}$ then scaling both $\Lambda$ and $\rho_m/c$ down so that  $\bar{\Omega}_{\Lambda}(t )$  is of order one and  $\bar{\Omega}_{M}(t )$ is negligible, to thus not require any cosmological dark matter.) The content of the fitting is  that even with  $\bar{\Omega}_{M}(t )$ being suppressed  so that $\bar{\Omega}_{\Lambda}(t )+\bar{\Omega}_{k}(t ) =1$, it could have been the case that $\bar{\Omega}_{\Lambda}(t )$ is huge and positive while  $\bar{\Omega}_{k}(t )$  is huge and negative (or vice versa).   Completely independent measurements (galactic rotation curves for $k$ and the accelerating universe data for $\bar{\Omega}_{\Lambda}(t )$) show that this is not the case. Thus by coupling with $G_{\rm eff}$ rather than with the Newtonian $G_{{\rm N}}$ conformal gravity self-quenches the contribution of the cosmological constant to cosmic evolution.

With $\tanh^2(\sigma^{1/2}ct_0)$ being fitted to $0.37$, we determine $\tanh(\sigma^{1/2}ct_0)=0.61$, $\sinh(\sigma^{1/2}ct_0)=0.77$, $\cosh(\sigma^{1/2}ct_0)=1.26$. To determine $a(t_0)$ (from which we can determine $a(t_R)$ as needed for the proper angular diameter distance) we need to determine $\sigma$. This is given from a determination of $H_0$ since $H(t_0)=\sigma^{1/2}c\cosh(\sigma^{1/2}ct_0)/\sinh(\sigma^{1/2}ct_0)$. This yields  $\sigma^{1/2}=0.47\times 10^{-28}$ ${\rm cm}^{-1}$. Then from $\sinh(\sigma^{1/2}ct_0)=0.77=\sinh(0.71)$ it follows that the current age of the universe is the perfectly acceptable $t_0=5.00\times 10^{17}$ sec. From the measured value of the spatial curvature $k$ it follows that $a(t_0)=2.51\times10^{-2}$ \cite{footnote2}. Consequently, with $a(t)T(t)$ taken to be  constant and $T_R/T_0=10^3$, $T_R=3\times 10^{3\circ}$K, we can set $a(t_R)=2.51\times10^{-5}$. For the recombination time we note that $\sinh(\sigma^{1/2}ct_R)=\sinh(\sigma^{1/2}ct_0)T_0/T_R=0.77\times 10^{-3}$. This value is small enough that we can set $\sigma^{1/2}ct_R=0.77\times 10^{-3}$. Thus $t_R=0.54\times 10^{15}$ sec. With this value for $t_R$ we obtain $\bar{\Omega}_{\Lambda}(t_R)= {\rm tanh}^2 (\sigma^{1/2}ct_R)=0.59\times 10^{-6}$, $\bar{\Omega}_k(t_R)= {\rm sech}^2 (\sigma^{1/2}ct_R)=1.00$, with the conformal cosmology recombination era being curvature dominated to one part in $10^6$.

We now have all the numerical data that we need to determine $d(\theta)$ and $D(t_F, t_R)$, and to start we need to determine the magnitude of $r_R$. Thus using (\ref{3.1}) we obtain 
\begin{eqnarray}
\int_0^{r_R} \frac{dr}{(1-kr^2)^{1/2}}&=&(-k)^{-1/2}\log[(-k)^{1/2}r_R+(1-kr_R^2)^{1/2}]
\nonumber\\
=\int_{t_R}^{t_0}\frac{c dt}{a(t)}&=&(-k)^{-1/2}\left[ \log\left(\frac{\sinh(\sigma^{1/2}ct_0)}{\cosh(\sigma^{1/2}ct_0)+1}\right)-\log\left(\frac{\sinh(\sigma^{1/2}ct_R)}{\cosh(\sigma^{1/2}ct_R)+1}\right)\right]
\nonumber\\
&=&(-k)^{1/2} \left[ \log\left(\frac{0.77}{2.26}\right)-\log\left(\frac{0.77\times 10^{-3}}{2}\right)\right]=
(-k)^{-1/2}\log\left(\frac{10^3}{1.13}\right).
\label{3.17}
\end{eqnarray}
From (\ref{1.2}) it follows that $(-k)^{1/2}=\gamma_0/2=1.53\times 10^{-30}$ cm. Thus for $r_R$ we obtain 
\begin{eqnarray}
r_R=(-k)^{-1/2}\frac{1}{2}\left[\frac{10^3}{1.13}-\frac{1.13}{10^3}\right]=6.54\times 10^{29}\times 442.48=
2.89\times 10^{32} {\rm cm}.
\label{3.18}
\end{eqnarray}
As we see, $r_R$ is indeed of cosmological size, with a magnitude that is fixed by the scalar three-curvature of the geometry, with this  three-curvature itself being determinable from a study of galactic rotation curves. 

To determine $d(\pi)$ we use (\ref{3.2}) and (\ref{3.18}) to obtain
\begin{eqnarray}
d(\pi)&=&2a(t_R)\int_0^{r_R} \frac{dr}{(1-kr^2)^{1/2}}=2a(t_R)(-k)^{-1/2}\log\left(884.96\right)
\nonumber\\
&=&2\times 2.51\times 10^{-5}\times 6.54\times 10^{29}\times 6.786~{\rm cm}=2.23\times 10^{26}~{\rm cm}.
\label{3.19}
\end{eqnarray}
From (\ref{1.2}) it follows that $\kappa^{-1/2}=3.24\times 10^{26}$ cm $\sim 105$ Mpc. We thus establish that the scale of $d(\pi)$ is given by none other than the inhomogeneous cluster scale that we determined from galactic rotation curves.

From (\ref{3.2}) the  arbitrary  $d(\theta)$ is given by
\begin{eqnarray}
d(\theta)=2a(t_R)(-k)^{-1/2}\log [(-k)^{1/2}r_R\sin(\theta/2)+(1-kr_R^2\sin^2(\theta/2))^{1/2}].
\label{3.20}
\end{eqnarray}
For $\theta=1^{\circ}=1/57.296$ radians (viz. of order the standard model horizon size) we have $(-k)^{1/2}r_R\theta/2=(-k)^{1/2}r_R/114.59=3.86$, and $\log [(-k)^{1/2}r_R\sin(\theta/2)+(1-kr_R^2\sin^2(\theta/2))^{1/2}]=\log(7.85)=2.06$. Thus
\begin{eqnarray}
d(1^{\circ})=2\times 2.51\times 10^{-5}\times 6.54\times 10^{29}\times 2.06=6.76\times 10^{25}~{\rm cm}.
\label{3.21}
\end{eqnarray}
 As we see, $d(1^{\circ})$ is also reasonably close to the inhomogeneous $\kappa^{-1/2}$ scale, so the $1^{\circ}$  region can also be associated with the $\kappa^{-1/2}$ scale. The $d(1^{\circ})/d(\pi)$ ratio is equal to $0.30$, so there is very little change in $d(\theta)$  across the recombination sky. This should be contrasted with the standard model picture, since, as we discuss in the simplified model analysis that we present in the appendix, with $k=0$ (\ref{3.2}) leads to the much smaller $d(1^{\circ})/d(\pi)=1^{\circ}/2=1/114.59=0.0087$.

 In order to evaluate  $D(t_F,t_R)$  and  $D(t^S_F,t_R,c_S)$ we note that since we can approximate $\sinh(\sigma^{1/2}ct_R)$ by $\sigma^{1/2}ct_R$, we can do so for any earlier but post very early universe time as well, so that for all such times we can set $a(t)=(-k)^{1/2}ct$. Thus 
 use of (\ref{3.3}) yields 
 \begin{eqnarray}
 D(t_F,t_R)=a(t_R)(-k)^{-1/2}\log\left(\frac{t_R}{t_F}\right)=a(t_R)(-k)^{-1/2}\log\left(\frac{T_F}{T_R}\right), \qquad D(t^S_F,t_R,c_S)=\frac{D(t_F,t_R)}{\sqrt{3}}.
 \label{3.22}
 \end{eqnarray}
Then with 
 \begin{eqnarray}
 d(\pi)=2a(t_R)(-k)^{-1/2}\log\left(\frac{10^3}{1.13}\right)
 \label{3.23}
 \end{eqnarray}
 the metric $D(t_F,t_R)$ and the acoustic $D(t^S_F,t_R,c_S)$ will be equal to $d(\pi)$ for  a $T_F$  and $T_F^S$ that obey 
 \begin{eqnarray}
 T_F=T_R\times \frac{10^6}{1.28}\approx 2.34\times 10^{9\circ}{\rm K}, \qquad T^S_F=T_R\times \left(\frac{10^6}{1.28}\right)^{\sqrt{3}}\approx 4.83\times 10^{13\circ}{\rm K}.
 \label{3.24}
 \end{eqnarray}

That there is a solution for $T_F$ at all means that conformal cosmology is causally connected, with there being no horizon problem \cite{footnote3}. That this solution occurs well below $T_V=10^{15\circ}$K means that it is not sensitive to any contribution of the matter fluid to cosmic evolution. Finally, and  most significantly, the obtained value for $T_F$ is in the primordial nucleosynthesis era, viz. an era of central importance to cosmology.  Our analysis here revises an earlier analysis of Mannheim and Horne that was reported in \cite{Mannheim2003}. Thus unlike in the standard cosmology where fluctuations start in the very early inflationary universe era, in conformal cosmology  gravitational fluctuations only need to start in the nucleosynthesis era in order to be able to fill out  the entire CMB recombination sky. Similarly, acoustic fluctuations only need to start at $10^{13\circ}{\rm K}$, i.e., after the mass generating phase transition at $10^{15\circ}{\rm K}$.

\section{Fluctuation amplitude growth}
\label{S4}
As well as discuss the rate of growth of the length scale we can also discuss the rate of growth of fluctuation amplitudes, as developed  in the  conformal cosmological fluctuation theory presented in \cite{Mannheim2012b,Amarasinghe2019,Phelps2019,Mannheim2020,Amarasinghe2021a,Amarasinghe2021b}. Central to doing this is the use of the scalar, vector, tensor (SVT) formalism presented in  \cite{Lifshitz1946,Bardeen1980} (and also in e.g. \cite{Kodama1984,Mukhanov1992,Stewart1990,Ma1995,Bertschinger1996,Zaldarriaga1998} and the reviews of \cite{Dodelson2003,Mukhanov2005,Weinberg2008,Lyth2009,Ellis2012}). In conformal time (viz. $c\tau=\int cdt/a(t)$) the background and fluctuation line element is given by
\begin{align}
ds^2&=-(g_{\mu\nu}+h_{\mu\nu})dx^{\mu}dx^{\nu}=\Omega^2(\tau)\left[c^2d\tau^2-\frac{dr^2}{1-kr^2}-r^2d\theta^2-r^2\sin^2\theta d\phi^2\right]
\nonumber\\
&+\Omega^2(\tau)\left[2\phi c^2d\tau^2 -2(\tilde{\nabla}_i B +B_i)cd\tau dx^i - [-2\psi\tilde{\gamma}_{ij} +2\tilde{\nabla}_i\tilde{\nabla}_j E + \tilde{\nabla}_i E_j + \tilde{\nabla}_j E_i + 2E_{ij}]dx^i dx^j\right].
\label{4.1}
\end{align}
In (\ref{4.1})  $\tilde{\nabla}_i=\partial/\partial x^i$ and  $\tilde{\nabla}^i=\tilde{\gamma}^{ij}\tilde{\nabla}_j$  (with Latin indices) are defined with respect to the background three-space metric $\tilde{\gamma}_{ij}$, and $(1,2,3)=(r,\theta,\phi)$. And with
\begin{eqnarray}
\tilde{\gamma}^{ij}\tilde{\nabla}_j V_i=\tilde{\gamma}^{ij}[\partial_j V_i-\tilde{\Gamma}^{k}_{ij}V_k]
\label{4.2}
\end{eqnarray}
for any three-vector $V_i$ in a three-space with three-space connection $\tilde{\Gamma}^{k}_{ij}$, the elements of (\ref{4.1}) are required to obey
\begin{eqnarray}
\tilde{\gamma}^{ij}\tilde{\nabla}_j B_i = 0,\quad \tilde{\gamma}^{ij}\tilde{\nabla}_j E_i = 0, \quad E_{ij}=E_{ji},\quad \tilde{\gamma}^{jk}\tilde{\nabla}_kE_{ij} = 0, \quad\tilde{\gamma}^{ij}E_{ij} = 0.
\label{4.3}
\end{eqnarray}
With the  three-space sector of the background geometry being maximally three-symmetric, it is described by a Riemann tensor of the form
\begin{eqnarray}
\tilde{R}_{ijk\ell}=k[\tilde{\gamma}_{jk}\tilde{\gamma}_{i\ell}-\tilde{\gamma}_{ik}\tilde{\gamma}_{j\ell}],
\label{4.4}
\end{eqnarray}
with general spatial three-curvature $k$. 
As written, (\ref{4.1}) contains ten fluctuation elements, whose transformations are defined with respect to the background spatial sector as four three-dimensional scalars ($\phi$, $B$, $\psi$, $E$) each with one degree of freedom, two transverse three-dimensional vectors ($B_i$, $E_i$) each with two independent degrees of freedom, and one symmetric three-dimensional transverse-traceless tensor ($E_{ij}$) with two degrees of freedom. The great utility of this basis is that since the cosmological fluctuation equations are gauge invariant, only gauge-invariant scalar, vector, or tensor combinations of the components of the scalar, vector, tensor basis can appear in the fluctuation equations. 
In \cite{Amarasinghe2021a} it was shown that for the fluctuations associated with the line element given in (\ref{4.1}) and with $\Omega(\tau)$ being an arbitrary function of $\tau$ and with $k$  being arbitrary, the gauge-invariant metric combinations are 
\begin{align}
\alpha=\phi + \psi + \dot B - \ddot E,\quad  \gamma= - \dot\Omega^{-1}\Omega \psi + B - \dot E,\quad  B_i-\dot{E}_i,  \quad E_{ij},
\label{4.5}
\end{align}
where the dot denotes $\partial/\partial c\tau$. In (\ref{4.5}) there is a total of six (one plus one plus two plus two) degrees of freedom, just as required since one can make four coordinate transformations on the initial ten components of the fluctuation $h_{\mu\nu}$. Interestingly, we note that even with $k\neq 0$ these gauge invariant  combinations have no explicit dependence on $k$.

While this analysis is general for any general covariant theory, in the conformal case there is also conformal invariance (arbitrary rescaling of the conformal factor $\Omega(\tau)$). This reduces the number of degrees of freedom by one, with the one gauge invariant scalar combination in (\ref{4.5}) that explicitly depends on $\Omega(\tau)$ (viz. $\gamma$) dropping out. And indeed, as shown in \cite{Amarasinghe2019}, only the $\alpha$, $B_i-\dot{E}_i$ and $E_{ij}$ combinations appear in the gravitational sector fluctuation $\delta W_{\mu\nu}$. On perturbing the matter sector $T_{\mu\nu}$ that appears in (\ref{3.7}), then as shown in \cite{Mannheim2020,Amarasinghe2021b} by working with gauge invariant combinations the  fluctuation equation $4\alpha_g\delta W_{\mu\nu}=\delta T_{\mu\nu}$ can be solved exactly without any approximation at all.

For our purposes here though we note that  while there are matter fluctuations in the matter sector perfect fluid, since this fluid plays no role at temperatures below $10^{15\circ}$K, in (\ref{3.7}) we only need to perturb 
\begin{equation}
T^{\mu \nu} = 
-\frac{1}{6}S_0^2\left(R^{\mu\nu}
-\frac{1}{2}g^{\mu\nu}R^\alpha_{\phantom{\alpha}\alpha}\right)         
-g^{\mu\nu}\lambda S_0^4,
\label{4.6}
\end{equation}                                 
to study the metric fluctuations associated with (\ref{4.5}) at temperatures below $10^{15\circ}$K. When this is done very straightforward conformal time perturbative fluctuation equations for $\alpha$, $B_i-\dot{E}_i$ and $E_{ij}$ are then obtained \cite{Mannheim2020,Amarasinghe2021b}:  
\begin{align}
\eta\Omega^{-2}[\ddot{\alpha}-2\dot{\Omega}\Omega^{-1}\dot{\alpha}-\tilde{\nabla}_b\tilde{\nabla}^b\alpha]&=-\alpha,
\nonumber\\
\eta\Omega^{-2}(\tilde{\nabla}_a\tilde{\nabla}^a-\partial_{\tau}^2-2k)(B_i-\dot{E}_i)
&=(B_i-\dot{E}_i),
 \nonumber\\
 \eta\Omega^{-2}\left[ (\tilde\nabla_b \tilde\nabla^b-\partial_{\tau}^2-2k)^2+4k\partial_{\tau}^2 \right] E_{ij}
&=- \ddot{E}_{ij} - 2 k E_{ij} - 2 \dot{E}_{ij} \dot{\Omega} \Omega^{-1} +\tilde{\nabla}_{a}\tilde{\nabla}^{a}E_{ij},
\label{4.7}
\end{align}
where $\eta=-24\alpha_g/S_0^2$. With $\alpha_g$ being known to be negative \cite{Mannheim2016}, $\eta$ is positive. In (\ref{4.7}) we note that the left-hand side of each relation is due to $\delta W_{\mu\nu}$ and the right-hand side is due to $\delta T_{\mu\nu}$. The key consequence of not including matter fluid perturbations is that the fluctuations are then driven by gravity itself, with the matter fluid acoustic oscillations then only responding to the metric fluctuations rather than influencing them. Like photon mode fluctuations,  gravitational metric fluctuations are fluctuations around the background light cone. While matter fluctuations could play a bigger role if we break the conformal symmetry  dynamically via $\langle \Omega|\bar{\psi}\psi|\Omega \rangle$ rather than the scalar field $S(x)$ \cite{Mannheim2017,Amarasinghe2021b}, we will not consider matter fluctuations here. This will enable us to establish what metric fluctuations can do on their own. 

Given  that $a(t)=(-k/\sigma)^{1/2}\sinh(\sigma^{1/2}ct)$ as per (\ref{3.14}) we obtain \cite{Amarasinghe2021b} 
\begin{eqnarray}
\rho=(-k)^{1/2}c\tau=\ln \tanh\left(\frac{\sigma^{1/2}ct}{2}\right),\qquad \sinh\rho=-\frac{1}{\sinh (\sigma^{1/2}ct)},\qquad \Omega(\rho)=-
\left(\frac{-k}{\sigma}\right)^{1/2}\frac{1}{\sinh\rho}.
\label{4.8}
\end{eqnarray}
(As defined $\rho$ is negative, and with $\rho(t=0)=-\infty$, $\rho(t=\infty)=0$, $\rho(t)$ decreases in magnitude as $t$ increases.) As discussed in  \cite{Mannheim2020,Amarasinghe2021b}, on setting $r=(-k)^{-1/2}\sinh\chi$ and then  $\alpha(\rho,\chi,\theta,\phi)=\alpha(\rho) S(\chi,\theta,\phi)$, the scalar field spatial sector (and analogously the vector and tensor sectors) can be separated according to $[\tilde{\nabla}_b\tilde{\nabla}^b+(-k)(\tau^2+1)]S(\chi,\theta,\phi)=0$, where $(-k)(\tau^2+1)$ is a separation constant and $\tau$ is a continuous variable that lies in the range $0\leq \tau \leq \infty$. Then, on setting $\varkappa=\sigma\eta=48\lambda \alpha_g$,  the conformal time dependence of $\alpha(\rho)$ is given by
\begin{align}
\left(\frac{d^2}{d\rho^2}+2\frac{\cosh\rho}{\sinh\rho}\frac{d}{d \rho}+\tau^2+1+\frac{1}{\varkappa \sinh^2\rho}\right)\alpha(\rho)=0,
\label{4.9}
\end{align}
with $\varkappa$ being positive since both $\sigma$ and $\eta$ are positive. Solutions to (\ref{4.9}) can be given in terms of associated Legendre conical functions of the first kind. Consequently, $\alpha(\rho)$ and its vector and tensor counterparts are given by \cite{Amarasinghe2021b}
\begin{align}
\alpha(\rho)&=\sinh^{-1/2}\rho P^{-1/2-K}_{-1/2+i\tau}(\cosh \rho),
\nonumber\\
B_i(\rho)-\dot{E}_i(\rho)&=\epsilon_i \sinh^{1/2}\rho P^{-1/2-K}_{-1/2+i\tau}(\cosh \rho),
\nonumber\\
E_{ij}(\rho)&=B_{ij} \sinh^{3/2}\rho P^{-1/2-K}_{-1/2+i\tau}(\cosh \rho),
\label{4.10}
\end{align}
where $\epsilon_i$ is a transverse polarization vector, where $B_{ij}$ is a transverse-traceless polarization tensor,  
and where  $K=-1/2+(1/4-1/\varkappa)^{1/2}$. (We could also use the second kind of conical function  $Q^{-1/2-K}_{-1/2+i\tau}(\cosh \rho)$, but for our purposes here the consequences are the same.) With this definition and with $K$ being taken to be real, $K$ has to lie between $-1/2$ and $0$, as fixed by $\varkappa$ lying between $4$ and $\infty$.  Setting $\varkappa$ equal to infinity is equivalent to setting $\eta=\infty$ in (\ref{4.7}). Thus $K=0$ represents the pure $\delta W_{\mu\nu}$ contribution to the fluctuations. The larger the magnitude of $K$ the larger the contribution of the matter sector $\delta T_{\mu\nu}$.
While the $P^{-1/2-K}_{-1/2+i\tau}(\cosh \rho)$ (or $Q^{-1/2-K}_{-1/2+i\tau}(\cosh \rho)$) are not generally known in closed form, we note that when $K=0$ (viz. 
$\varkappa=\infty$),  the  first and second kind of conical functions are of the form
\begin{align}
P^{-1/2}_{-1/2+i\tau}(\cosh \rho)= 
\frac{1}{(2\pi\sinh\rho)^{1/2}}\frac{(e^{i\tau\rho}-e^{-i\tau\rho})}{i\tau}, \qquad
Q^{-1/2}_{-1/2+i\tau}(\cosh \rho)=- \left(\frac{\pi}{2\sinh\rho}\right)^{1/2}\frac{e^{-i\tau\rho}}{\tau},
\label{4.11}
\end{align}
with both having the same large $\rho$ behavior.

On setting $S(\chi,\theta,\phi)=S_{\ell}(\chi)Y^m_{\ell}(\theta,\phi)$ the scalar sector radial dependence is given by 
\begin{eqnarray}
\left[\frac{d^2}{d\chi^2}+2\frac{\cosh\chi }{\sinh\chi}\frac{d }{ d\chi}
-\frac{\ell(\ell+1)}{ \sinh^2\chi}+(\tau^2+1)\right]S_{\ell}(\chi)=0.
\label{4.12}
\end{eqnarray}
While (\ref{4.12}) is also an associated Legendre conical function equation, since $\ell$ is integer its solutions are known in a closed form  \cite{Bander1966}, \cite{Phelps2019,Mannheim2020, Amarasinghe2021b}, viz.
\begin{eqnarray}
S_{\ell}(\chi)&=&\sinh^{\ell}\chi\left(\frac{1}{ \sinh\chi} \frac{d }{ d\chi}\right)^{\ell+1}\cos\tau\chi,\qquad
N(\tau,\ell)=\left(\frac{2}{\pi\tau^2(\tau^2+1^2)....(\tau^2+\ell^2)}\right)^{1/2},
\label{4.13}
\end{eqnarray}
as normalized with the normalization factor $N(\tau,\ell)$. The first few $S_{\ell}(\chi)$  are of the form
\begin{align}
S_0&=-\frac{\tau\sin(\tau\chi)}{\sinh\chi}, \qquad S_1=-\frac{\tau^2\cos(\tau\chi)}{\sinh\chi} +\frac{\tau\sin(\tau\chi)\cosh\chi}{\sinh^2\chi},
\nonumber\\
S_2&=\frac{\tau^3\sin(\tau\chi)}{\sinh\chi}-\frac{2\tau\sin(\tau\chi)}{\sinh\chi}
+\frac{3\tau^2\cos(\tau\chi)\cosh\chi}{\sinh^2\chi}-\frac{3\tau\sin(\tau\chi)}{\sinh^3\chi}.
\label{4.14}
\end{align}
As noted in  \cite{Phelps2019,Mannheim2020,Amarasinghe2021b} the spatial sectors of the $B_i-\dot{E}_i$ and $E_{ij}$ fluctuations obey completely analogous associated Legendre function equations, with $B_i-\dot{E}_i$ behaving as $S_{\ell}(\chi)/\sinh\chi$ and $E_{ij}$ behaving as $S_{\ell}(\chi)/\sinh^2\chi$.

With $dr^2/(1-kr^2)$ being equal to $(-k)^{-1}d\chi^2$, radial modes on the light cone obey $d\rho^2-d\chi^2=0$. With $\chi$ necessarily being positive since $r=(-k)^{-1/2}\sinh\chi$ is positive, and with $\rho$ as constructed in (\ref{4.8}) being negative, unperturbed light cone modes travel on $\chi=-\rho$ trajectories. With our definition of $\rho$ and $\chi$, both of them will be large in magnitude when the comoving time $t$ is small, and both will be small in magnitude when $t$ is large. As discussed in \cite{Phelps2019,Mannheim2020}, without needing to  set $\chi=-\rho$ the scalar fluctuation $\alpha$ given in (\ref{4.5}) behaves as $e^{-\chi}e^{-|\rho |}$ at large $\chi$ and $|\rho|$, and as $\chi^{\ell}|\rho|^{K}$ at small $\chi$ and $|\rho|$.  For the vector fluctuation $B_{\chi}-\dot{E}_{\chi}$ the analogous limits are $e^{-2\chi}e^{0}$ and $\chi^{\ell-1}|\rho|^{K+1}$. For the tensor fluctuation $E_{\chi\chi}$ the analogous limits are $e^{-3\chi}e^{|\rho |}$ and $\chi^{\ell-2}|\rho|^{K+2}$.  With $\chi=|\rho|$ all three fluctuation share the same two limits, viz. $e^{-2\chi}$ and $\chi^{\ell +K}$. We note that for the scalar fluctuations the minimum allowed $\ell$ is $\ell=0$, for the vector fluctuations the minimum allowed $\ell$ is $\ell=1$, while for the tensor fluctuations the minimum allowed $\ell$ is $\ell=2$ \cite{Phelps2019,Mannheim2020}. While we can thus take $\alpha$ to be representative of all the fluctuations, we note that since $K$  has to lie in the $-1/2\leq K\leq 0$ range, the $\ell=0$ scalar modes with nonzero $K$ would blow up at $\chi=0$. (It is for this reason that we have chosen not to set $\rho$, and thus $\chi$, to be zero at the current time $t_0$, but rather in (\ref{4.8}) we have set $\rho$ to vanish at $t=\infty$ \cite{footnote5}.)

As had been noted above, at recombination $\sinh (\sigma^{1/2}ct_R)$ is quite small, being equal to $0.77\times 10^{-3}$. Thus from (\ref{4.8}) we see that $\sinh\rho_R$ is large. Moreover, given the pattern exhibited in (\ref{4.14}), for any $\ell$ the leading behavior of $S_{\ell}(\chi)$ is of order $1/\sinh \chi$ at large $\sinh\chi$. With $\sinh\rho_R$ being large, so is $\cosh\rho_R$. For large $\cosh\rho$ the asymptotic behavior of $P^{-1/2-K}_{-1/2+i\tau}(\cosh \rho)$ is  known to be of the form
\begin{align}
P^{-1/2-K}_{-1/2+i\tau}(\cosh \rho)&\rightarrow \frac{1}{\cosh^{1/2}\rho}\left[\frac{\Gamma(i\tau)e^{i\tau\rho}}{(2\pi)^{1/2}\Gamma(i\tau+K+1)}+\frac{\Gamma(-i\tau)e^{-i\tau\rho}}{(2\pi)^{1/2}\Gamma(-i\tau+K+1)}\right]
\label{4.15}
\end{align}
for any $K$ and $\tau$.
Thus for any $K$ or $\tau$ the large $\rho$ behavior is of the form $(\cosh\rho)^{-1/2}$, i.e., of the form $(\sinh\rho)^{-1/2}$, just as shown in the exact $K=0$ case given in (\ref{4.11}). Combining (\ref{4.13}) and (\ref{4.15}) we see that any $\tau,\ell,m$ mode of $\alpha(\rho,\chi,\theta,\phi)$ is given by
\begin{align}
\alpha(\rho,\chi,\theta,\phi)&=\frac{N(\tau,\ell)}{\sinh\rho\sinh\chi}\left(\frac{d^{\ell+1}\cos(\tau\chi)}{d \chi} \right)\left[\frac{\Gamma(i\tau)e^{i\tau\rho}}{(2\pi)^{1/2}\Gamma(i\tau+K+1)}+\frac{\Gamma(-i\tau)e^{-i\tau\rho}}{(2\pi)^{1/2}\Gamma(-i\tau+K+1)}\right]Y^m_{\ell}(\theta,\phi)
\label{4.16}
\end{align}
at large $\sinh\rho$ and $\sinh\chi$. Now for modes fluctuating about the light cone we have $\chi=-\rho$. 
Thus at recombination the full $\alpha(\rho_R,\chi_R=-\rho_R)$ behaves as $1/\sinh^2\rho_R$, i.e., as $\sinh^2(\sigma^{1/2}ct_R)$. Since this same behavior holds at any earlier time ($\sinh(\sigma^{1/2}ct)$ being even smaller), comparing recombination with any earlier time we obtain
\begin{eqnarray}
\frac{\alpha(\rho_R,\chi_R=-\rho_R)}{\alpha(\rho,\chi=-\rho)}=\frac{\sinh^2(\sigma^{1/2}ct_R)}{\sinh^2(\sigma^{1/2}ct)}=\frac{a^2(t_R)}{a^2(t)}=\frac{T^2(t)}{T^2_R}
\label{4.17}
\end{eqnarray}
as the rate of amplitude growth for a mode with any given $\ell$.
This is a quite substantial growth rate. For instance, in going from nucleosynthesis to recombination the fluctuation amplitude grows by a factor of $(10^9/10^3)^2=10^{12}$. Moreover, since $\sinh(\sigma^{1/2}ct)$ does not become big even at the present time, in going from recombination to the current era the fluctuation amplitude grows by a factor of order $10^{6}$. While technically we cannot apply this analysis above $10^{15}$ degrees, if we do so just to get an order of magnitude estimate, we see that a proton-sized radius of order $10^{-13}$ cm would grow to the recombination $d(\pi)=2.23\times 10^{26}$ cm if it started out at a temperature of order $10^{23}$ degrees at a time of order $10^{-5}$ seconds. Similarly, we note that a length as small as $10^{-33}$ cm would grow to be as big as the recombination $d(\pi)$ if it started out a temperature of order $10^{33}$ degrees at a time of order $10^{-15}$ seconds, being able to do so without any period of exponential expansion.

We should note that while $\alpha(\rho,\chi, \theta, \phi)$ remains small all the way to the current era, it does increase with comoving time since $a(t)$ does grow as the universe expands, while remaining smaller than one until the current era. Thus perturbation theory is valid for $\alpha(\rho,\chi, \theta, \phi))$ all the way to $t_0$.  While modes with any $\ell$ all grow at the same rate  in time, to determine whether the conformal theory can fit the angular momentum maxima and minima seen in the anisotropy of the CMB requires comparing the amplitudes  of modes with differing $\ell$ at the same recombination time.  Work on this in conjunction with T. Liu and D. A. Norman is currently in progress.

\section{Summary}
\label{S5}

Unlike the standard local Newton-Einstein gravitational theory the conformal gravity theory is a global theory, with distant matter sources having a significant effect on the dynamics within local systems such as galaxies. Consequently,  study of the motion of test particles in galaxies provides information not only on the local gravitational field within any given galaxy but also on the global gravitational field due to the rest of the material in the universe. From studies of the systematics of the rotation curves of spiral galaxies two global effects have been identified, one being due to the background global cosmology and the other being due to inhomogeneities in it. The background cosmology provides a scale $\gamma_0=3.06\times 10^{-30}~{\rm cm}^{-1}$, while the inhomogeneities provide a scale $\kappa=9.54\times10^{-54}~{\rm cm}^{-2}$, viz. $\kappa^{-1/2}=3.24\times 10^{26}~{\rm cm}=105$ Mpc. In this paper we have shown that the $\kappa^{-1/2}$ scale is imprinted on the recombination CMB sky, with the proper angular diameter distances  $d(\pi)$ and $d(1^{\circ})$ being of this very magnitude. We note that a similar such sized  inhomogeneous scale is associated with both baryon acoustic oscillations and the cosmic web. In addition, in this paper we have shown that the fluctuations that are imprinted on and able to fill out the recombination  CMB sky only need to originate in the nucleosynthesis era and not at any altogether earlier juncture. The length scale expansion rate is thus far more rapid than in the standard cosmology. As well as study the growth rate of the metric length scale we have also studied the growth rate of the metric fluctuation amplitude, and found that  it grows between nucleosynthesis and recombination by a factor of $10^{12}$.

\begin{acknowledgments}
The author wishes to acknowledge a very fruitful collaboration with Dr. K. D. Horne, and very helpful discussions with T. Liu and D. A. Norman.
\end{acknowledgments}

\appendix
\numberwithin{equation}{section}
\setcounter{equation}{0}

\section{Simplified treatment}
\label{SA}
For the exact solution for $a(t)$ as given (\ref{3.12}), viz.
\begin{eqnarray}
a^2(t)= -\frac{k(\beta-1)}{2\sigma}
-\frac{k\beta{\rm sinh}^2 (\sigma^{1/2} ct)}{\sigma},
\label{A.1}
\end{eqnarray}
the integral $\int cdt/a(t)$ can actually be done analytically. It is of the form
\begin{eqnarray}
(-k)^{1/2}\int_0^t\frac{cdt}{a(t)}=\beta^{-1/2}
F({\rm arcsin}({\rm sech}(\sigma^{1/2}ct)),(\beta+1)^{1/2}/(2\beta)^{1/2})-\beta^{-1/2}F(\pi/2,(\beta+1)^{1/2}/(2\beta)^{1/2}),
\label{A.2}
\end{eqnarray}
viz. an incomplete elliptical integral of the first kind  with $j^{\prime 2}=(\beta-1)/2\beta=1-j^2$, $j^2=(\beta+1)/2\beta$ in the standard elliptical integral notation with $F(\phi,j)=\int_0^{\phi}dx/(1-j^2\sin^2x)^{1/2}$, $F({\rm arcsin}({\rm sech}x),j)=\int dx/(j^{\prime 2}+\sinh^2x)^{1/2}$. We note that $F(\pi/2,1)=\infty$, as would have to be the case since when $\beta=1$ $a(t)$ behaves as $(-k)^{1/2}ct$ with $\int dt/t$ then diverging at $t=0$. Thus for studies in the very early universe where the $\beta$-dependent term is important we must keep $\beta=(1- 16A\lambda/k^2c)^{1/2}$ not equal to one so as to enable $\int dt /a(t)$ to not be singular at $t=0$. 

While we would have to use (\ref{A.2}) for an exact treatment, a simplified treatment can be obtained by noting that with the largest value for $\sinh(\sigma^{1/2}ct)$  in the $0\leq t \leq t_0$ range  being at $t_0$, then since $\sinh(\sigma^{1/2}ct_0)=0.77$ is reasonably small, other than in  the very early universe we can approximate $a(t)$ by
\begin{eqnarray}
a(t)=\frac{ct}{L},
\label{A.3}
\end{eqnarray}
where we have set $k=-1/L^2$ \cite{footnote6}. Then with $a(t_0)/a(t_R)=t_0/t_R=T_R/T_0$, (\ref{3.1}), (\ref{3.2}) and (\ref{3.3}) give
\begin{align}
L\log\left[\frac{r_R}{L}+\left(1+\frac{r_R^2}{L^2}\right)^{1/2}\right]&=L\log\left(\frac{t_0}{t_R}\right)
=L\log\left(\frac{T_R}{T_0}\right),
\nonumber\\
d(\theta)&=2ct_R\log\left[\frac{r_R\sin(\theta/2)}{L}+\left(1+\frac{r_R^2\sin^2(\theta/2)}{L^2}\right)^{1/2}\right],
\nonumber\\
D(t_F,t_R)&=ct_R\log\left(\frac{T_F}{T_R}\right),\qquad D(t^S_F,t_R,c_S)=ct_R\log\left(\frac{T_F}{T_R}\right)^{\sqrt{3}}.
\label{A.4}
\end{align}
With $T_R \gg T_0$, it follows that
\begin{eqnarray}
\frac{2r_R}{L}=\frac{T_R}{T_0},
\label{A.5}
\end{eqnarray}
\begin{eqnarray}
d(\pi)=2ct_R\log\left(\frac{T_R}{T_0}\right)=2\times 3\times 10^{10}\times 0.54 \times 10^{15}\times 6.908=2.24\times 10^{26}~{\rm cm},
\label{A.6}
\end{eqnarray}
in agreement with (\ref{3.19}). Then, on setting the metric $D(t_F,t_R)=d(\pi)$, and the acoustic  $D(t^S_F,t_R,c_S)=d(\pi)$ we find that  $T_F$ and $T_F^S$ are given by 
\begin{eqnarray}
T_F=\frac{T_R^3}{T_0^2}=3\times 10^{9\circ}{\rm K},\quad T_F=T_R\left(\frac{T_R^2}{T_0^2}\right)^{\sqrt{3}}=7.40\times 10^{13\circ}{\rm K},
\label{A.7}
\end{eqnarray}
in general agreement  with (\ref{3.24}). 

When $\theta$ is small  we can approximate $d(\theta)$ by 
\begin{eqnarray}
d(\theta)=2ct_R\log\left[\frac{T_R\theta}{4T_0}+\left(1+\frac{T_R^2\theta^2}{16T_0^2}\right)^{1/2}\right].
\label{A.8}
\end{eqnarray}
There are now three limits we could consider: $T_R\theta/4T_0 \gg1$,  $T_R\theta/4T_0 =1$ and $T_R\theta/4T_0 \ll 1$. The three limits respectively lead to 
\begin{align}
d(T_R\theta/4T_0 \gg1) &=2ct_R\log\left(\frac{T_R\theta}{2T_0}\right),
\nonumber\\
d(T_R\theta/4T_0 =1) &=2ct_R\log(1+\sqrt{2}]=2\times 0.88 ct_R=2.85 \times10^{25} ~{\rm cm},
\nonumber\\
d(T_R\theta/4T_0 \ll1) &=ct_R\frac{T_R\theta}{2T_0}.
\label{A.9}
\end{align}
For $\theta=1^{\circ}$   we have $T_R\theta/4T_0 =10^3/(4\times 57.30)=4.36$, so the first limit applies approximately, and we obtain
\begin{eqnarray}
d(1^{\circ}) =2\times 3\times 10^{10}\times 0.54\times 10^{15}\times 2.17= 7.03\times 10^{25}~{\rm cm},
\label{A.10}
\end{eqnarray}
which agrees with (\ref{3.21}) \cite{footnote7}. For $\theta=1/4.36=0.23^{\circ}$ the second limit applies, while for $\theta =0.1^{\circ}$ we have $T_R\theta/4T_0 =.436$, so  the third limit applies approximately, and we obtain
\begin{eqnarray}
d(0.1^{\circ}) =3\times 10^{10}\times 0.54\times 10^{15}\times .872= 1.41\times 10^{25}~{\rm cm}.
\label{A.11}
\end{eqnarray}

In the $k=0$ standard model we have  $\int _0^{r_L\sin(\theta/2)}dr=r_L\sin\theta/2$. Thus for small $\theta$ we can set
$r_L\sin\theta/2=r_L\theta/2$ no matter how big or small $r_L$ might be. Thus in the standard model we have $d(1^{\circ})/d(\pi)=1^{\circ}/2=0.0087$. Consequently, the standard model $1^{\circ}$ proper angular diameter distance is substantially smaller than the standard model full CMB sky $d(\pi)$.  However, in the conformal gravity case where $k$ is negative we need to compare $\theta$ with $T_R/T_0$. Thus we can have small $\theta$ and yet have large $T_R\theta/4T_0$. In this case we obtain
\begin{eqnarray}
d(T_R\theta/4T_0 \gg1) =2ct_R\log\left(\frac{T_R\theta}{2T_0}\right)=d(\pi) +2ct_R\log\left(\frac{\theta}{2}\right).
\label{A.12}
\end{eqnarray}
As we see, while $d(\theta)$ is reduced from $d(\pi)$ since $\theta/2$ is less than one radian if $\theta$ is small, it is only reduced by a logarithmic factor and thus not by the big factor reduction that occurs in the standard model. Thus in the conformal theory $d(\theta)$ does not vary radically across the recombination CMB sky, with $d(1^{\circ})/d(\pi)=0.31$. 

For the amplitude of the fluctuations it is more direct to first write the conformal time $\alpha(\rho)$ equation associated with (\ref{4.7}) in comoving time $t$, viz. \cite{Mannheim2020,Amarasinghe2021b}
\begin{align}
\eta\left(\frac{a^2}{c^2}\frac{d^2\alpha}{dt^2}-\frac{a}{c^2}\frac{da}{dt}\frac{d\alpha}{dt}+\frac{(\tau^2+1)}{L^2}\alpha\right)+ a^2\alpha=0.
\label{A,13}
\end{align}
On using (\ref{A.3}) for $a(t)$, then, as noted in \cite{Mannheim2020},  we obtain 
\begin{eqnarray}
\left[\frac{\eta}{c^2}\left(\left(\frac{\partial}{\partial t}\right)^2 -\frac{1}{t}\frac{\partial}{\partial t}+\frac{\tau^2+1}{t^2}\right)+1\right]\alpha=0,\quad 
\left[\frac{\eta}{c^2}\left(\left(\frac{\partial}{\partial t}\right)^2 +\frac{1}{t}\frac{\partial}{\partial t}+\frac{\tau^2}{t^2}\right)+1\right]\left(\frac{\alpha}{ct}\right)=0.
\label{A.14}
\end{eqnarray}
The solution to the second equation in (\ref{A.14}) is a  Bessel function, and on excluding the irregular Bessel function since it is badly behaved at $t=0$, the solution to (\ref{A.14}) that is real is given in comoving time by
\begin{align}
&\alpha=ct[A(\tau)J_{i\tau}(\eta^{-1/2}ct)+A^*(\tau)J_{-i\tau}(\eta^{-1/2}ct)],
\label{A.15}
\end{align}
where $A(\tau)$ is a constant. For small $\eta^{-1/2}ct$  we find that $\alpha(t)$ behaves as
\begin{align}
&\alpha\rightarrow ct\left[A(\tau)\left(\frac{\eta^{-1/2}ct}{2}\right)^{i\tau}+A^*(\tau)\left(\frac{\eta^{-1/2}ct}{2}\right)^{-i\tau}\right].
\label{A.16}
\end{align}

Now in the solution $a(t)=(-k/\sigma)^{1/2}\sinh(\sigma^{1/2}ct)$ for  which (\ref{A.3}) is an approximation,  $\sinh\rho$ is given in (\ref{4.8}) as $\sinh\rho=-1/\sinh(\sigma^{1/2}ct)$. Thus we can set
\begin{align}
e^{\rho}=\frac{\cosh(\sigma^{1/2}ct)-1}{\sinh(\sigma^{1/2}ct)},\qquad e^{-\rho}=\frac{\cosh(\sigma^{1/2}ct)+1}{\sinh(\sigma^{1/2}ct)},
\label{A.17}
\end{align}
with small $\sigma^{1/2}ct$ behavior 
\begin{align}
e^{\rho}=\frac{\sigma^{1/2}ct}{2},\qquad e^{-\rho}=\frac{2}{\sigma^{1/2}ct}.
\label{A.18}
\end{align}
Now, according to (\ref{4.10}) and (\ref{4.15}) the large $\rho$ behavior of $\alpha(\rho)$ is given by
\begin{align}
\alpha(\rho)\rightarrow  \frac{1}{\sinh\rho}\left[\frac{\Gamma(i\tau)e^{i\tau\rho}}{(2\pi)^{1/2}\Gamma(i\tau+K+1)}+\frac{\Gamma(-i\tau)e^{-i\tau\rho}}{(2\pi)^{1/2}\Gamma(-i\tau+K+1)}\right].
\label{A.19}
\end{align}
With (\ref{A.18}) we recognize (\ref{A.19}) as (\ref{A.16})  with an appropriate choice for $A(\tau)$.

To convert to conformal time we have $L\rho=\int cdt/a(t)=L[\ln(ct)+{\rm constant}]$. We shall take the constant to be $\ln(\sigma^{1/2}/2)$ so that $\rho=\ln(\sigma^{1/2}ct/2)$. Then from (\ref{A.15}) we obtain 
\begin{align}
\alpha
\rightarrow 2\sigma^{-1/2}e^{\rho} \left[A(\tau)\left(\frac{e^{\rho}}{\sigma^{1/2}\eta^{1/2}}\right)^{i\tau}+A^*(\tau)\left(\frac{e^{\rho}}{\sigma^{1/2}\eta^{1/2}}\right)^{-i\tau}\right],
\label{A.20}
\end{align}
with $\sigma\eta=48\lambda\alpha_g$. Now $\rho$ is negative. Thus as $\rho \rightarrow -\infty$ we have $1/\sinh\rho=2/(e^{\rho}-e^{-\rho}) \rightarrow -2e^{\rho}$. Thus we can rewrite (\ref{A.20}) as 
\begin{align}
\alpha
\rightarrow -\frac{1}{\sigma^{1/2}\sinh\rho} \left[A(\tau)\left(\frac{e^{\rho}}{\sigma^{1/2}\eta^{1/2}}\right)^{i\tau}+A^*(\tau)\left(\frac{e^{\rho}}{\sigma^{1/2}\eta^{1/2}}\right)^{-i\tau}\right].
\label{A.21}
\end{align}
We thus recover (\ref{A.19}) and identify $A(\tau)$ as 
\begin{align}
A(\tau)=-\frac{\sigma^{1/2}(\sigma\eta)^{i\tau/2}\Gamma(i\tau)}{(2\pi)^{1/2}\Gamma(i\tau+K+1)}.
\label{A.22}
\end{align}
\numberwithin{equation}{section}
\setcounter{equation}{0}


\begin{thebibliography}{99}
\bibitem{footnote2} In the $k=0$ case it is conventional to normalize so that $a(t_0)=1$. With $k<0$ we  do not use this normalization here.
\bibitem{footnote4} If one sets $a(t)=t^n$ then $\int dt/a(t)=t^{1-n}/(1-n)$, to thus diverge at $t=0$ if $n \geq 1$, and be finite if $n<1 $ (cf. the standard model $t^{1/2}$ radiation era).  Hence for $n\geq 1$ the whole universe is contained within one and the same causal domain. Now while inflation also puts the universe into a single causal domain, it does so with an exponential expansion rate, a rate that is much larger than a $t^n$ power behavior. It thus gets outside the $t^{1/2}$  expansion rate horizon at $1^{\circ}$K  by more than is needed, and  leads to a lot of power in the angular scale temperature fluctuation power spectrum at large angles. While beyond the scope of this paper it would be of interest to see whether conformal gravity with its power-behaved (not exponential) expansion rate might not lead to  this amount of power at large angles.
\bibitem{footnote5} To set $\rho(t_0)=-\chi(t_0)=0$ we  introduce an integration constant in $\int cdt/a(t)$ so that $\rho(t)=-\chi(t)=\log[\sinh(\sigma^{1/2}ct)/(\cosh(\sigma^{1/2}ct)+1)]-\log[\sinh(\sigma^{1/2}ct_0)/(\cosh(\sigma^{1/2}ct_0)+1)]$.
Then $\rho(t_R)=-\chi(t_R)=-7.86+1.08=-6.78$.
\bibitem{footnote6} As such, a linear in $t$ expansion rate is the temporal analog of the linear in $r$ behavior given in (\ref{2.8}).
\bibitem{footnote7}  In \cite{Mannheim2003} it was incorrectly thought that  in analog to the standard gravity treatment, $d(1^{\circ})$ should be associated with the $T_R\theta/4T_0 \ll 1$  limit rather than with the $T_R\theta/4T_0 \gg 1$ limit. In the conformal gravity case it is  $d(0.1^{\circ})$ that can be associated with the $T_R\theta/4T_0 \ll 1$  limit.
\end{thebibliography}
\end{document}